# Measurement of a Phase of a Radio Wave Reflected from Rock Salt and Ice Irradiated by an Electron Beam for Detection of Ultra-High-Energy Neutrinos

Masami Chiba[1], Toshio Kamijo[1], Takahiro Tanikawa[1], Hiroyuki Yano[1], Fumiaki Yabuki[1], Osamu Yasuda[1], Yuichi Chikashige[2], Tadashi Kon[2], Yutaka Shimizu[2], Souichirou Watanabe[2], Michiaki Utsumi[3], Masatoshi Fujii[4]

[1]*Graduate School of Science and Engineering, Tokyo Metropolitan University, 1-1 Minami-Ohsawa, Hachioji-shi, Tokyo 192-0397, Japan*
[2]*Faculty of Science and Technology, Seikei University, Musashino-shi, Tokyo 180-8633, Japan*
[3]*Department of Applied Science and Energy Engineering, School of Engineering, Tokai University, Hiratsuka-shi, Kanagawa 259-1292, Japan*
[4]*School of Medicine, Shimane University, Izumo-shi, Shimane 693-8501, Japan*

We have found a radio-wave-reflection effect in rock salt for the detection of ultra-high energy neutrinos (UHEν's) which are expected to be generated in Greisen, Zatsepin, and Kuzmin (GZK) processes in the universe. When an UHEν interacts with rock salt or ice as a detection medium, a shower is generated. That shower is formed by hadronic and electromagnetic avalanche processes. The energy of the UHEν shower converts to thermal energy through ionization processes. Consequently, the temperature rises along the shower produced by the UHEν. The refractive index of the medium rises with temperature. The irregularity of the refractive index in the medium leads to a reflection of radio waves. This reflection effect combined with the long attenuation length of radio waves in rock salt and ice would yield a new method to detect UHEν's. We measured the phase of the reflected radio wave under irradiation with an electron beam on ice and rock salt powder. The measured phase showed excellent consistence with the power reflection fraction which was measured directly. A model taking into account the temperature change explained the phase and the amplitude of the reflected wave. Therefore the reflection mechanism was confirmed. The power reflection fraction was compared with that calculated with the Fresnel equations, the ratio between the measured result and that obtained with the Fresnel equations in ice was larger than that of rock salt.

## I. INTRODUCTION

Ultra-high-energy neutrinos (UHEν's) were predicted to be produced at a collision of UHE-cosmic ray with the cosmic-microwave-radiation background by Berezinsky and Zatsepin [1] in Greisen, Zatsepin, and Kuzmin (GZK) processes in the universe [2, 3]. In order to detect them we had searched for and found a radio-wave-reflection effect in rock salt [4 – 6]. A gigantic detector is needed for the detection due to the expected ultra-low flux of about 1 km$^{-2}$ · d$^{-1}$. When an UHEν interacts with rock salt or ice as a detection medium, a shower is generated. That shower is formed by hadronic and electromagnetic avalanche processes. The energy of the UHEν shower converts finally to thermal energy through ionization processes. Consequently, the temperature rises along the shower produced by the UHEν. The refractive index rises as a function of the temperature. The irregularity of the refractive index in the medium for radio waves causes reflection. This reflection effect combined with long attenuation length of radio waves in rock salt and ice would yield a new method to detect UHEν's. We could find a huge amount of rock salt or ice of over 50 Gt in natural rock salt formations or the Antarctic ice sheet. The volume of the rock salt is 3 × 3 × 3 km$^3$ which is required for the detection of the neutrino flux. Radio waves transmitted into the medium generated by a radar system with a phased array antenna could be reflected by the shower. Receiving the reflected radio waves could be a method for the detection of the UHEν's.

We had carried out an experiment [4] previously to the one described in this article to observe microwave reflection effects from a small rock-salt sample of $2 \times 2 \times 10$ mm$^3$ irradiated by a synchrotron radiation X-ray with a pulse width of 1.7 s. It was set in a 9.4 GHz waveguide [4] while microwaves were continuously injected to the waveguide. A null-detection method was employed to detect the feeble reflected signal in the waveguide circuit. Reflected microwaves were observed with a power reflection fraction of $1 \times 10^{-6}$ and a decay time of 8 s after the irradiation had stopped. The shape of the power reflection fraction with respect to time was similar to that of the temperature change in rock salt. The reflection fraction was proportional to the square of the X-ray intensity.

A larger rock salt sample of 10 cm$^3$ was irradiated by a 2 MeV electron beam with a duration of 60 s which was set in a free space without using a waveguide [5]. A continuous 435 MHz radio wave struck at the cube from a six-element Uda-Yagi antenna. The reflected power fraction increased as the temperature rose at the irradiated surface of the cube. The observation of the reflection excluded the possibility that the effect was due to a distortion of the waveguide heated by X-ray irradiation at the experiment [4].

A 435 MHz waveguide filled with rock salt powder was used to measure the radio wave reflection effect [6]. The 2 MeV electron beam was injected to rock salt powder through an aluminum-beam window of $20 \times 20$ cm$^2$ with a thickness of 0.5 mm. A continuous 435 MHz radio wave of $10^{-4}$ W was emitted by a quarter-wavelength antenna installed in the salt powder and the reflected radio wave was detected by the same antenna. The reflection fraction increased with the square of the temperature rise at the irradiated surface of the rock salt powder. The reflection fraction could be explained by the Fresnel equations.

## II. EXPERIMENT

We report now on a new experiment using a coaxial tube (WX-20D) with a diameter of 20 mm and a length of 100 mm. An electron beam was injected into an open end of the coaxial tube set in a dry-ice cooling box. The coaxial tube was filled with rock salt powder or ice and was set in the cooling box as shown in Fig. 1. The temperature was measured by a chromel-alumel thermocouple positioned 2 mm from the surface of the open end. The refractive indices of rock salt powder and ice were measured by the reflection method from 0.4 MHz to 1 GHz in the coaxial tube yielding 1.79 $\pm$ 0.01 at 22 ℃ and 1.76 $\pm$ 0.01 at -60 ℃, respectively. Then the diameter of the inner conductor was set to 4.5 mm in order to obtain an impedance of 50 Ω when the tube was filled with the respective medium.

We utilized a 2 MeV electron beam produced by a Cockcroft-Walton accelerator located at the Takasaki Advanced Radiation Research Institute (TARRI) of Japan Atomic Energy Agency (JAEA). The electron beam of 2 MeV was irradiated on the open end of the coaxial tube with a power of 4.2 J/s and a current of 1 mA. For the irradiation of a large target and to prevent damage of a titanium vacuum-beam window, the electron beam was swept over the target with a width of 1 m at 200Hz. Only a small part of the beam hit the open end of the coaxial tube. The increase of the refractive index with respect to the temperature gave rise to radio-wave reflection. We observed the reflection effect from ice as well as rock salt.

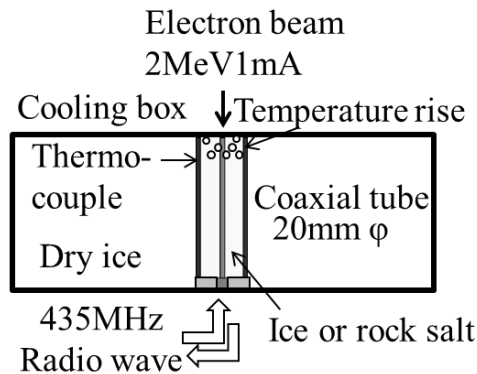

**FIGURE 1.** A coaxial tube with a diameter of 20 mm and a length of 100 mm was filled with rock salt power or ice. It was put in a cooling box of $300 \times 300 \times 100$ mm$^3$. The temperature in the medium was measured 2 mm under the open surface. An electron beam was injected into the open end.

As shown in Fig. 2, a 435 MHz continuous wave of $10^{-4}$ W from an Oscillator (Rohde & Schwarz SMB100A) was split (Mini-Circuits ZMSC-2) into a signal $\vec{a}$ (expressed as a vector) sent to the coaxial tube through a circulator (MTC B115FFF) and reference signal $\vec{d}$ that could be phase shifted by a variable phase shifter (Mini-Circuits JSPHS-446). The reflected signal $\vec{b}$ from the coaxial tube was sent to a combiner (Mini-Circuits ZMSC-2) through the circulator. The combined signal of $\vec{c} = \vec{b} + \vec{d}$ was split into a real-time spectrum analyzer (Tektronix RSA3303B) and a detector (Power Detector, Mini-Circuits ZX47-60-S+) for the power measurement. In order to reject noises due to the power source of 50 Hz, from the real-time spectrum analyzer (RSA), 1024 data samples within 128 ms in the time domain were fast-Fourier transformed to the frequency domain. The peak at 435 MHz was selected within $\pm$ 8 Hz to reject noise. In order to tune the frequency precisely between the oscillator and RSA, they were locked by the 10

MHz signal generated by a Rubidium-frequency standard (Stanford Research Systems FS725).

The output signal of the detector was routed to an Educational-Laboratory-Virtual-Instruments Suite (National Instruments NI ELVIS) based on an NI LabVIEW system where the signal was digitized by an ADC.

The digital output was converted to a value, such that the output of the combiner became zero, by a control program of LabVIEW and converted to an analog voltage by a DAC in the NI ELVIS. The analog signal was fed to the variable phase shifter as a control voltage where the phase of the reference signal $\vec{d}$ was shifted according to the voltage. The output of the phase shifter was fed to the combiner again. The feedback loop was repeated at ~ 13 Hz.

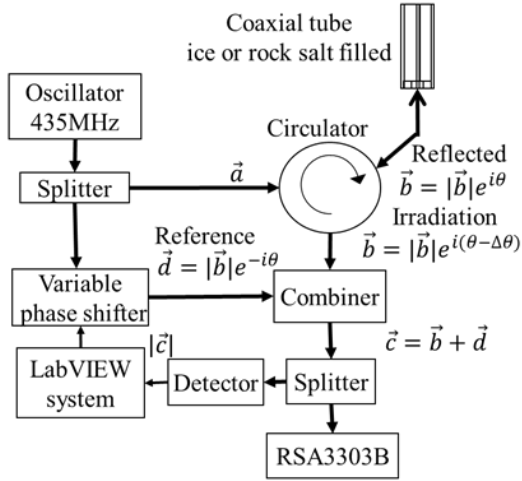

**FIGURE 2.** A 435MHz continuous wave of $10^{-4}$ W from an Oscillator (Rohde & Schwarz SMB100A) was split (Mini-Circuits ZMSC-2) into a signal $\vec{a}$ sent to the coaxial tube through a circulator (MTC B115FFF) and another as a reference signal $\vec{d}$ that was phase shifted by a variable phase shifter (Mini-Circuits JSPHS-446). The reflected signal $\vec{b}$ from the coaxial tube was sent to a combiner (Mini-Circuits ZMSC-2) through the circulator. The combined energy was measured by a real-time spectrum analyzer (Tektronix RSA3303B) and an ADC in NI ELVIS based on NI LabVIEW system through a detector, respectively. An output of the DAC in the NI ELVIS was fed to a variable phase shifter (Mini-Circuits JSPHS-446) to minimize $|\vec{c}|$, repeating a feedback in a loop.

A measurement of the power reflection fraction was done as follows. Before the electron beam irradiation, the vector of reflection signal was $\vec{b} = |\vec{b}|e^{i\theta}$ where a phase of θ was a constant without the irradiation. The phase of the reference signal was tuned to $-\theta$ so that the output of the combiner $\vec{c} = \vec{b} + \vec{d}$ became close to zero. The amplitude of the reference signal was tuned slightly by a variable attenuator (Mini-Circuits ZX73-2500-s) which was set just after the variable phase shifter so as to realize $\vec{d} = |\vec{b}|e^{-i\theta}$. By simultaneously tuning the variable phase shifter and the variable attenuator in the feedback loop, the proper amplitude and phase of $\vec{d}$ was maintained. The control voltages of the variable phase shifter and the attenuator were recorded in each loop to know the phase shift and the attenuation. When the irradiation began, we stopped the feedback loop through the control program i.e. $\vec{d} = |\vec{b}|e^{-i\theta}$ was fixed. At the same time, the phase Δθ of the reflection signal began to change in $\vec{b} = |\vec{b}|e^{i(\theta-\Delta\theta)}$. Consequently, the reflection combined amplitude of $|\vec{c}| = |\vec{b} + \vec{d}|$ began to increase and was measured by the RSA and the ADC of NI ELVIS.

In case of a measurement of the phase, we did not stop the feedback loop when the irradiation started. The phase was tuned automatically in the loop to get $|\vec{c}|$ close to zero in the same way as the amplitude measurement. The reference signal was maintained as $\vec{d} = |\vec{b}|e^{i(-\theta+\Delta\theta)}$. From the recorded phase of $-\theta + \Delta\theta$, we deduced Δθ of the reflected signal.

Figure 3 shows a result of the measurements for an ice target. The electron beam of 2 MeV with 2 mA was irradiating the target for 60 s. As the beam irradiated, the temperature increased from -60 ℃ to -35 ℃ over the time. The temperature was measured by the aforementioned thermocouple which was recorded by a NI CompactDAQ. The power reflection fraction measured by the RSA is plotted as small closed circles which is the reflection power fraction between the reflected wave and the wave injected to the coaxial tube. Five data points which were obtained by sampling over an interval of 128 ms are in 1 s. They increased from 0 to $8 \times 10^{-7}$ as the temperature rose and traced a curve without large fluctuations. The phase Δθ obtained from the variable phase shifter decreased from 0° to -0.2°. The relative refraction indices of $n$ in ice [7] and the rock salt powder [8] increased with the temperature as in equations (1) and (2), respectively, where $T$ is expressed in degrees Celsius.

$$n = 0.000260T + 1.786 \quad (1)$$

$$n = 0.000498T + 1.781 \quad (2)$$

Because the density of the rock salt powder is less than that of rock salt, the refractive index of rock salt powder is smaller than that of rock salt. We measured the refractive index of the rock salt powder and ice by a reflection method in a coaxial tube. The decrease of the phase in the measurement is explained by the decrease of the velocity of the wave in the ice due to the increase of $n$.

The power of the reflected waves was calculated by Eq. (3) from the resulting vector $\vec{c}$ of a vector subtraction between two vectors with the length of $|\vec{b}|$ and the rotation angle of $\Delta\theta$:

$$|\vec{c}|^2 = 2|\vec{b}|^2 \{1 - cos(\Delta\theta)\} \quad (3)$$

The power reflection fraction was obtained by a calculation using the phase difference $\Delta\theta$ and plotted as a short dash in Fig. 3. They are somewhat scattered but coincide very well with the RSA data.

A model calculation was done to confirm the cause of the radio wave reflection effect. We assumed that the temperature of the ice from the open end to 10 mm within the coaxial tube was the same as the measured temperature at 2 mm from the open end. We calculated the power reflection fraction based on the telegrapher's equations. The result was in agreement with the RSA value within 50 % during the irradiation and the shape was similar to the curve describing the temperature as a function of time.

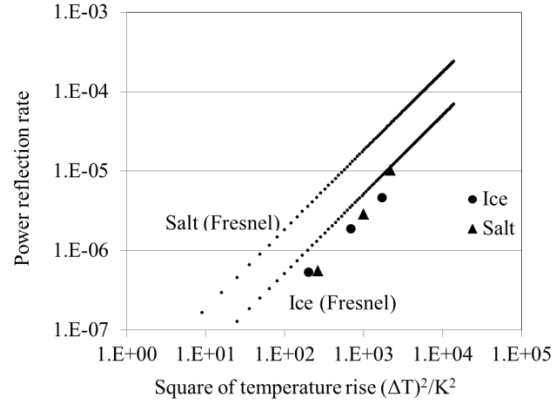

**FIGURE 4.** Power reflection fraction with respective to the square of the temperature rise are shown. Calculated values and measured values of the salt powder or the ice are depicted as "Salt (Fresnel)" or "Ice(Fresnel)" by a line connected by dots and "Salt" by closed triangles or "Ice" by closed circles, respectively.

$$\Gamma = \frac{(n_2 - n_1)^2}{(n_2 + n_1)^2} \quad (4)$$

The values of $\Gamma$, calculated by Eq. (4) using Eq. (1) and the measured values of the salt powder are depicted as "Salt (Fresnel)" by a line connected by dots and the measured value "Salt" by closed triangles, respectively. A line connected by dots of "Ice (Fresnel)" is also calculated by Eq. (4). As can be seen, "Salt (Fresnel)" is 3.6 times larger than "Ice (Fresnel)", but the experimental data of "Ice" of closed circles and "Salt" of closed triangles are roughly the same with respect to the square of the temperature rise. The data of "Salt" and "Ice" are 16 and 54 % compared with "Salt (Fresnel)" and "Ice (Fresnel)", respectively. The loss of the reflection fraction was partly due to imperfect signal transmission of the coaxial tube and the partial measurement of the temperature. The higher power reflection fraction in "Ice" compared to "Salt" suggests that the ice was partly melted along an electron track locally. The refractive index of water at 0 ℃ is 5.3 times larger than that of ice. It might enhance the power reflection fraction.

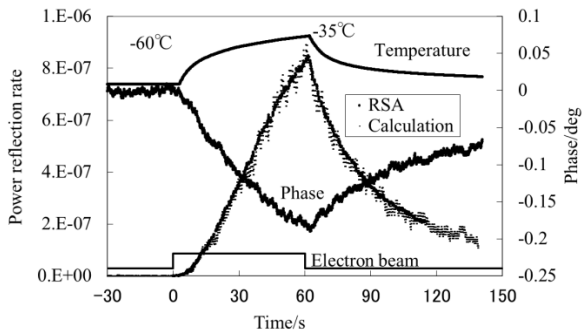

**FIGURE 3.** During irradiation with the electron beam, the temperature of the ice target increased from -60 ℃ to -35 ℃ with time. The power reflection fraction measured by the RSA is plotted as small closed circles and increased from 0 to $8 \times 10^{-7}$ with the temperature. The phase $\Delta\theta$ obtained from by the variable phase shifter decreased from 0° to -0.2°. The power reflection fraction was got by a calculation using the phase $\Delta\theta$ and plotted as a short dash.

The power reflection fraction with respect to the square of the temperature rise is shown in Fig. 4. We compared the power reflection fractions at the irradiation time of 60 s for the beam currents of 1, 2 and 3 mA with the power reflection fraction $\Gamma$ of Fresnel equations as shown in Eq. (4). The refractive indices $n_1$ and $n_2$ are calculated from the measured temperatures before and during the irradiation, respectively.

## III. SUMMARY

We found a radio wave reflection effect from ice as well as rock salt through irradiation with an electron beam. In addition to the amplitude measurement, we measured the phase of the reflected radio wave from ice and rock salt by constructing an automatic feedback loop for the null detection method in which the variable phase shifter was tuned to get the null output. The fraction of the reflected power was calculated from the difference of two vectors with the same length and rotated with respect to each other by the rotation angle. Moreover we explained the reflection by a model in which the phase delay of the reflected wave was caused by the increase of the refractive index due to the temperature rise. The radio wave reflection effect is applicable to a radar method to detect GZKν in a huge amount of rock salt formation, Antarctic ice sheet and the moon crust.

Five types of radiation detectors using thermal effects are compared in Table 1. A "Radar Chamber" as well as a "Bolometer" detects a temperature rise due to an energy deposition of incident radiation. It receives the reflected radio wave (10 MHz – 1 GHz) from a heated portion emitted by a transmitter. It is appropriate to detect weakly interacting particle like neutrinos since the medium is a solid with a large density. The change of the refractive index of a medium is rather small due to the smaller temperature rise in the large volume of the shower. However the coherent reflection effect between the reflected waves from the many heated points distributed in the large volume of the shower might enhance the reflection fraction for wavelengths that are long compared to the shower diameter. Even with a coherence effect the reflected radio wave is weak and a "Radar Chamber" is suitable only for a large energy deposit. The detector size depends on the reflection fraction and the attenuation length of radio waves in the medium. The attenuation along the path could be compensated by a strong and narrow beam of radio waves generated artificially. We could detect GZKν's using radio waves with a peak power of 1 GW (Equivalent Isotropic Radiation Power) where the radio waves are supplied by a phased array antenna set on a surface of the medium. We do not need expensive boreholes for the installation of antennas. A "Radar Chamber" has a long memory time and we could scan the effective volume by a radio wave beam within the memory time. So it could be operated as a stand-alone detector without a trigger detector system.

According to the elucidation of the radio wave reflection mechanism, a new radiation detector "Radar Chamber" is applicable for all dielectric media where the refractive indices change with temperature. It is not only for radiation detection but also for other purposes using materials with inhomogeneous refractive index in space and time. An application for human-body imaging could be investigated at 10 MHz where the attenuation length is ~ 5 cm.

**TABLE 1.** Radiation detectors using heat effects.

|  | Bolometer | Cloud Chamber | Bubble Chamber | Acoustic Detector | Radar Chamber |
|---|---|---|---|---|---|
| Inventor | S.P. Langley | C.T.R. Wilson | D.A. Glaser | G.A. Askaryan | M. Chiba, et al. |
| Year | 1878 | 1911 | 1952 | 1957 | 2007 |
| Medium | Solid | Gas | Liquid | Solid, Liquid | Solid |
| Wave length | - | ~ 500 nm | ~ 500 nm | ~ 1m | 0.3 ~ 30m |
| Body | Absorber | Liquid particle, | Bubble, | Heated portion, | Heated portion, |
| Body size | - | ~ 0.5 mm | ~ 0.1 mm | $0.1 \varphi \times 5$ m | $0.1 \varphi \times 5$ m |
| Reflection or Emission | - | Reflection | Reflection | Emission | Reflection |
| Operation | - | Decompression | Decompression | - | - |
| Process | Heating | Super cooling | Super heating | Heating | Heating |
| Amplification | Small heat capacity | Growth of liquid particle | Growth of bubble | - | Coherent reflection |
| Sensitivity | > 1 eV | ~ 100 eV | ~ 100 eV | > $10^{12}$ eV | > $10^{12}$ eV |
| Position resolution | - | ~ 0.5 mm | ~ 0.1 mm | ~ 30 m | ~ 30 m |
| Detector size | ~ 1 cm | ~ 1 m | ~ 3 m | ~ 3 km | ~ 3 km |
| Memory time | ~ 1s | ~10 ms | ~ 1 μs | - | ~ 10 s |

## ACKNOWLEDGEMENTS

This work is partially supported by a Grant in Aid for Scientific Research for Ministry of Education, Science, Technology and Sports and Culture of Japan, and Funds of Tokutei Kenkyuhi and Tokubetsu Kenkyuhi at Tokyo Metropolitan University and Seikei University, respectively.

This work has been performed at the station of AR-NE5A, KEK under the approval of the Photon Factory Program Advisory Committee (Proposal No. 2004P009, 2005G121) and has been supported by the Inter-University Program for the Joint Use of JAEA Facilities.